\documentstyle[12pt,eqsecnum,aps,prd,preprint]{revtex}
\tightenlines
\input epsf.tex
\begin{document}
\preprint{\baselineskip=18pt \vbox{\hbox{SU-4240-669}\hbox{November 1997}}}
\title{\large\bf Remark on the potential function of the linear sigma model}
\author{David Delphenich and Joseph Schechter}
\address{Physics Department, Syracuse University\\ 
Syracuse, NY 13244-1130} 
\maketitle
%\receipt{November 10, 1997}
\begin{abstract}
It is shown that the potential functions for the ordinary linear sigma model can be divided into two topographically different types depending on whether the quantity $R\equiv(m_\sigma/m_\pi)^2$ is greater than or less than nine.  Since the Wigner-Weyl mode ($R=1$) and the Nambu-Goldstone mode ($R=\infty$) belong to different regions, we speculate that this classification may provide a generalization to the broken symmetry situation, which could be convenient for roughly characterizing different possible applications of the model.  It is noted that a more complicated potential does not so much change this picture as add different new regions.
\end{abstract}
\vskip 1cm
\section {Introduction}
The well-studied linear sigma model \cite{sigma} has been enormously useful as a source of physical insight into the phenomenon of spontaneous symmetry breaking in field theory.  Even at the tree level it provides a plausible simplified treatment of the low energy interactions of pions.  In the present note, we point out a possibly amusing description of the intrinsically different regions of the potential function in terms of the tree level ratio, R, of sigma to pion squared masses:
\begin{equation}
R=(\frac{m_\sigma}{m_\pi})^2.
\end{equation}

We have two motivations for this work.  First, it was noted many years ago \cite{Co} that the stable particles in the strong coupling limit of ``two flavor,'' two-dimensional QED \cite{Gepner} comprise a negative parity isotriplet (``pion'') and a positive parity isosinglet (``sigma'').  Their masses satisfy $R=3$.  Since this particle content agrees with that of the linear sigma model, it is natural to ask \cite{Us} if that model can give a low energy description of the theory and, if so, what significance should be attached to the particular value of R.  A second motivation arises from recent phenomenological theories of ordinary pion-pion scattering by several authors \cite{pipi,typical value of R} which require the existence of a low energy broad scalar resonance.  A typical value, \cite{pipi} $m(\sigma) \approx 550$ MeV, corresponds to $R \approx 16.$

In section II, we investigate in detail the usual fourth order potential functions of the linear sigma model.  The basic idea for classifying their types comes from ``catastrophe theory'' \cite{catastrophe}.  A small modification is made to take account of chiral invariance and the dependence of the boundary lines on $R$ is studied.  In section III, for the purpose of testing the sensitivity of the results to modification of the potential, we treat the sixth order potential functions.  Finally, section IV contains a brief summary and mention of directions for extending this work.

\section{Linear sigma model potential}

The fields of the linear sigma model \cite{sigma} consist of a negative parity isotriplet, $\pi_i, i=1, 2, 3,$ and a positive parity isosinglet, $\sigma.$  The Lagrangian density is
\begin{equation}
{\cal L} = -\frac{1}{2}(\partial_\mu \bbox{\pi})^2 - \frac{1}{2}
(\partial_\mu \sigma)^2 - V(\sigma, \bbox{\pi}),
\end{equation}
\begin{equation}
V(\sigma,\bbox{\pi})=A(\sigma^2 + \bbox{\pi}^2 - \lambda)^2 - B\sigma,
\end{equation}
where $A>0, B$ and $\lambda$ are three real constants.  One must take $A>0$, to enforce stability.  The constant $\lambda$ may have either sign.  For $\lambda > 0$ we have a ``wrong sign mass term'' which (taking $\pi_2= \pi_3=0$ for ease of visualization) leads to the famous ``Mexican hat'' shape when we set $B=0$ and plot $V$ against $\sigma$ and $\pi_1$ (Fig. 1a).  For $\lambda <0$ there is a ``correct sign mass term'' and, when we set $B=0$, $V(\sigma, \pi_1)$ has no local maximum (Fig. 1b).

Note that when $B=0$, the Lagrangian is invariant under chiral $SU(2)_L \times SU(2)_R$ transformations,
\begin{equation}
M \to U_L M U_R^\dagger,
\end{equation}
where $M=\frac{1}{\sqrt{2}}(\sigma + i\ \bbox{\tau}\cdot\bbox{\pi})$ and $U_L, U_R$ are two-dimensional unitary unimodular matrices.

We treat the model, for our present purpose, in the semi-classical approximation.  The theory is to be expanded around the global minimum ($\langle\sigma\rangle, \langle\pi_i\rangle$).  We shall restrict attention to the parity-conserving and isospin-conserving case \footnote{When we compare the minimization equation for $\pi_i: \langle\frac{\partial V}{\partial\pi_i}\rangle=4A\langle\pi_i\rangle (\langle\sigma\rangle^2+\langle\pi_i\rangle^2-\lambda)=0$ with the corresponding equation for $\sigma$ we learn that $\langle\pi_i\rangle\ne 0$ implies $B=0$.  In other words, if the chiral symmetry breaking coefficient is present we must have $\langle\pi_i\rangle=0$.  When $B=0$, a chiral rotation enables us to consistently also choose $\langle\pi_i\rangle=0$.},
 which satisfies
\begin{equation}
\langle\pi_i\rangle=0.
\end{equation}
It is necessary to impose the restriction to the global minimum of the potential:
\begin{equation}
\langle \frac{\partial V}{\partial \sigma}\rangle \equiv \frac{\partial V}{\partial \sigma} |_{\sigma=\langle\sigma\rangle, \pi_i=0}=4A\langle\sigma\rangle(\langle\sigma\rangle^2-\lambda)-B=0,
\end{equation}
The masses of the pions and $\sigma$ particle become:
\begin{equation}
m_\pi^2 = \langle\frac{\partial^2V}{\partial\pi_3\partial\pi_3}\rangle= 4A(\langle\sigma\rangle^2-\lambda)=\frac{B}{\langle\sigma\rangle}\ge 0,
\end{equation}
\begin{equation}
m_\sigma^2= \langle\frac{\partial^2V}{\partial\sigma^2}\rangle\
= 4A(3\langle\sigma\rangle^2-\lambda)\ge 0,
\end{equation}
The last equality in (2.6) occurs only for $\langle\sigma\rangle\ne0$. The ratio of the masses squared is evidently:
\begin{equation}
R= \frac{3\langle\sigma\rangle^2-\lambda}{\langle\sigma\rangle^2-\lambda}.
\end{equation} 
In this formula $\langle\sigma\rangle$ is the global minimum obtained from solving (2.5).

Of course, when $B=0$, there may be a continuous set of global minima as in Fig 1a.  This spontaneous breakdown situation occurs when $\lambda=\langle\sigma\rangle^2>0$ and is seen from (2.6) and (2.7) to result in $R=\infty$.  On the other hand, for $B=0$ and $\lambda<0$, we must have (from eq. (2.5)) $\langle\sigma\rangle=0$ and hence $R=1.$  Summarizing these well-known results for the symmetric Lagrangian case,
\begin{equation}
R=\frac{m^2_\sigma}{m^2_\pi}= \left\{\begin{array}{l}\infty, \qquad (\lambda>0,B=0) \\1, \qquad (\lambda<0,B=0). \end{array}\right.
\end{equation}

When $B\ne 0$, it is like adding a tilted plane to $V(\sigma, \pi_1)$.  In the case when a Mexican hat shape is present it gets tilted.  Instead of having a circle of degenerate minima, there is only one global minimum.  This is illustrated in Fig. 1c.

Our goal is to give an intrinsic characterization of the entire $(\lambda, A, B)$ space in terms of $R$.  Incidentally, the $\lambda=0$ case is seen from (2.8) to yield
\begin{equation}
R=3,\qquad (\lambda=0, |B|>0),
\end{equation}

Let us simplify the notation by setting
\begin{equation}
x\equiv\sigma, \qquad u\equiv-2\lambda, \qquad v\equiv-B/A,
\end{equation}
and defining the function
\begin{equation}
W(x)/A \equiv x^4 + ux^2 + vx
\end{equation}
Then
\begin{equation}
V(x,0) = W(x) + A\lambda^2.
\end{equation}

It is clear that since we are restricting attention to the parity-conserving $\langle\pi_i\rangle=0$ case and since $V(x,0)$ differs from $W(x)$ by a constant we can equivalently find $\langle x \rangle$ as a global minimum of $W(x)$.  This is helpful since the function in (2.12) has been extensively analyzed under the heading of ``catastrophe theory'' \cite{catastrophe}.  In this theory a topographical property of the potential (i.e., how many local minima it possesses) is studied as a function of the ``control parameters'' $u$ and $v$.  We also see that, for fixed $u$ and $v$, the behavior of the potential is independent of the positive quantity $A$.  Thus, it is sufficient to study the behavior of our potential as a function of $u$ and $v$.

Now, depending on the values of $u$ and $v, W(x)$ may have either one or two minima.  This follows since the critical point equation
\begin{equation}
W^{\prime}(x)/A = 4x^3+2ux+v=0,
\end{equation}
can have either one or three real roots.  The latter situation yields two minima and one maximum.  We want to find the locus of points in the $u-v$ plane which separates the region of one minimum from the region of two minima.  The transition occurs when a maximum and a minimum in the two minima region coalesce to form a point of inflection.  The condition for this is that the second derivative vanish:
\begin{equation}
W^{\prime\prime}(x)/A= 12x^2+2u=0
\end{equation}
The desired locus in the $u-v$ plane is obtained by eliminating $x$ between (2.14) and (2.15) to give \cite{catastrophe}:
\begin{equation}
u=-\frac{3}{2}v^{2/3}.
\end{equation}
This cusp is plotted in Fig. 2.  Inside the cusp (hatched region) there are two minima.  Outside the cusp, there is only one minimum.  Note that, unlike $W$, $V$ is a function of $\bbox\pi$ as well as $\sigma$.  This means that the two-minima region for $W$ translates into the region where $V$ possesses a local maximum with a surrounding valley (See Fig. 1a for the case where the valley is horizontal with a continuum of degenerate minima and Fig. 1c for the case where the valley is tilted with a single global minimum.  In this case, the other local minimum of $W$ becomes a saddle point.).  The one minimum region for $W$ translates simply to a region for $V$ with one minimum, but no local maximum.

Note that for the symmetric Lagrangian case, where $B=-vA=0$, the Mexican hat potential containing a local maximum belongs to the hatched region within the cusp while the chiral symmetric spectrum potential (with $\lambda<0$) belongs to the region outside the cusp.  Thus the lines of the cusp provide an intrinsic generalization to the case when the symmetry-breaker $B$ is present, of the well-known separation of the symmetric potentials into those of the Nambu-Goldstone type ($\lambda>0$) and those of the Wigner-Weyl type ($\lambda<0$).  The generalized Nambu-Goldstone type potentials have a local maximum (unstable critical point), in addition to a surrounding valley (usually tilted), while the generalized Wigner-Weyl type potentials just have a single global minimum.

Next, we point out that the cusp lines which divide the two regions in $u-v$ space are characterized by
\begin{equation}
R=9	\qquad (for\ u= -\frac{3}{2}v^{2/3}).
\end{equation}
To see this, first observe that since a maximum and a minimum coalesce for a potential with $u$ and $v$ parameters on the cusp, $W^{\prime}(x)$ must have a double root there:
\begin{equation}
W^{\prime}(x)/A =4(x-a)(x-b)^2.
\end{equation}
Equating this to (2.14) and comparing powers of $x$ yields
\begin{equation}
b=-\frac{a}{2},\qquad u=-\frac{3}{2}a^2,\qquad v=-a^3.
\end{equation}
Evidently the global minimum for cusp potentials is at $x=a.$  Hence, the desired ratio can be evaluated, using (2.6), (2.7) and (2.11) as
\begin{equation}
R=\frac{4Aa}{B}(3a^2-\lambda)=(\frac{4a}{-v})(3a^2+\frac{u}{2}).
\end{equation}
With the help of (2.19) we immediately get (2.17).

We summarize the $R$ values characterizing the axes and cusp (from (2.9), (2.10) and (2.17)) in Fig. 3.  As one travels in a circle about the origin starting from a point on the positive $u$ axis, the value of $R$ increases monotonically from 1 (where the theory is in the Wigner-Weyl mode) to $\infty$ on the negative $u$ axis (where the theory is in the Nambu-Goldstone mode).  The generalized Nambu-Goldstone region is entered when $R=9$.  We may say that the theory is in a $\it{generalized}$ Nambu-Goldstone mode for $R>9$ and in a $\it{generalized}$ Wigner-Weyl mode for $R<9.$  There are no theories with $0<R<1.$

It is natural to suggest that the value of $R$ for a given theory furnishes a convenient, though rough, criterion for the nature of the theory.  Let us now check this for the two possible applications of the $SU(2)$ linear sigma model mentioned in the introduction.

If we speculate \cite{Us} that the linear sigma model adequately describes at least the vacuum for two-flavor $QED_2$ at low energies then we have the result \cite{Co} that $R=3.$  According to the above criterion, the theory would be in a generalized Wigner-Weyl mode.  If we consider the zero fermion mass limit of the model $(B=0)$ the Mermin-Wagner theorem \cite{MW} tells us that spontaneous symmetry breakdown is ruled out.  Hence, the classification of the model as being within the ``region of influence'' of the Wigner-Weyl mode is very reasonable.

If we try to apply the linear sigma model to low energy QCD, the recent analysis \cite{pipi} of $\pi\pi$ scattering which gives $m(\sigma) \approx 550MeV$ leads to $R \approx 16$.  This would, according to our criterion, put low energy QCD in a generalized Nambu-Goldstone mode.  Of course, the success of chiral perturbation theory \cite{chiral} suggests that this is also very reasonable.

It should be recognized that the chiral perturbation theory approach, based on a nonlinear sigma model with many higher derivative terms, is a more accurate model of low energy QCD than the linear sigma model.  Introduction of the sigma field can be accomplished in a more general fashion by starting out with the nonlinear model of pions and treating the sigma as a ``matter field'' \cite{CCWZ}.  This was the approach employed in \cite{pipi}.  Considering this aspect, the fact that 16 is not very different from 9 might perhaps be interpreted as meaning that low energy QCD may be close to the boundary of being in the generalized Wigner-Weyl mode.  The physical significance of being in one or the other generalized modes is, of course, related to the starting point for making a perturbation expansion.  We have given a rough criterion, but a more precise criterion might best be given in a generalized chiral perturbation theory framework.  The Orsay group \cite{Orsay} has undertaken an extensive program to investigate a similar issue. 

To avoid confusion, we stress that our results are purely ``kinematical''.  Namely, we have shown that the regime where the potential of the broken linear sigma model has a global minimum and the regime where it has a global minimum plus a local maximum are separated by a boundary which has the simple characterization $R=\frac{m_\sigma^2}{m_\pi^2}=9.$  We have not discussed the exact calculable differences between these two regions except to notice that the Wigner-Weyl mode belongs to the first while the Nambu-Goldstone mode belongs to the second.  Furthermore, we have not included the undoubtedly important effects of going beyond tree level.  This topic has been revived recently in some interesting papers \cite{sigma plus}.  (Of course, we are not regarding the linear sigma model as a fundamental theory.)

Finally, it is of interest to ask: ``what are the parameters $u$ and $v$ (defined in (2.11)) which correspond to fitting low energy QCD to the linear sigma model?''  For this purpose we need the identification of the ``pion decay constant'' $F_\pi$ as
\begin{equation}
\sqrt{2}\langle\sigma\rangle = F_\pi \approx 0.131\  GeV,
\end{equation}
where we have taken $F_\pi$ to be positive in the last step.  Then the minimization condition (2.5) gives the following straight line in the $u-v$ plane:
\begin{equation}
v=-\sqrt{2} F_\pi(F^2_\pi+u).
\end{equation}
Note that the result $m^2_\pi=\frac{B}{\langle\sigma\rangle}$ implies that $\langle\sigma\rangle$ must be positive for $B$ positive, i.e., for $v$ negative.  Eq. (2.22) is plotted in Fig. 4 corresponding to negative $v$. Without specifying $m_\sigma$, the theory could be anywhere along the line (2.22).  If we set $R=(m_\sigma/m_\pi)^2=16$ we get $u=-1.49\times10^{-2}\ GeV^2$ using (2.8) with (2.19) and $v=-4.19\times 10^{-4}\ GeV^3$ from (2.22).  As mentioned, this point is inside the cusp.

\section{Sixth order potential}

It is clearly of interest to explore the sensitivity of the preceding results to the form of the potential.  Since the analysis is, in fact, intrinsic to the form of the potential we must, a priori, expect the results to be different.  We test this out now for a sixth order, rather than a fourth order, potential function.  Interestingly, it turns out that when the parameters are in the regimes where both potentials are qualitatively similar, the results are also similar.  The special value $R=9$ for the fourth order case yields, in a sense to be described, to the special value $R\approx 13.253$ for the sixth order case.  This number is related to the real root of a certain cubic equation.  The sixth order potential also contains a regime which is qualitatively different from the fourth order case.

So now consider replacing the fourth order potential in (2.1) by 
\begin{equation}
V(\sigma, \bbox{\pi})=A[{(\sigma^2+\bbox{\pi}^2)^3+t(\sigma^2 +\bbox{\pi}^2)^2+v(\sigma^2+\bbox{\pi}^2)+w\sigma}],
\end{equation}
where $A>0$, while $t$, $v$, and $w$ are some new real constants.  As before, we either have, or may set, $\langle\pi_i\rangle=0$.  Furthermore, the minimization equation reads
\begin{equation}
\langle\frac{\partial V}{\partial\sigma}\rangle=A[6\langle\sigma\rangle^5+4t\langle\sigma\rangle^3+2v\langle\sigma\rangle+w]=0.
\end{equation}
The $\sigma$ and $\bbox{\pi}$ squared masses are given by
\begin{equation}
m_\sigma^2=\langle\frac{\partial^2 V}{\partial\sigma^2}\rangle = A[30\langle\sigma\rangle^4+12t\langle\sigma\rangle^2+2v] \ge 0,
\end{equation}
\begin{equation}
m_\pi^2=\langle \frac{\partial^2 V}{\partial\pi_1\partial\pi_1}\rangle=A[6\langle\sigma\rangle^4+4t\langle\sigma\rangle^2+2v]=-\frac{wA}{\langle\sigma\rangle}\ge 0,
\end{equation}
where the last equality in (3.4) holds for $\langle\sigma\rangle\ne 0$.  For convenience, we take the ratio of (3.3) and (3.4) to get our desired object:
\begin{equation}
R=\frac{m^2_\sigma}{m^2_\pi}=\frac{15\langle\sigma\rangle^4+6t\langle\sigma\rangle^2+v}{3\langle\sigma\rangle^4+2t\langle\sigma\rangle^2+v}.
\end{equation}
The analysis of $V(\sigma,\bbox{\pi})$ is closely related to that of
\begin{equation}
W(x)=V(x,0)=A[x^6+tx^4+vx^2+wx].
\end{equation}
This is an example (\footnote{If we add a second symmetry-breaking term, $u\sigma(\sigma^2+\bbox{\pi}^2)$ to $V(\sigma,\bbox{\pi})$, the resulting $W(x)$ is of a standard form \cite{catastrophe} which is ``structurally stable'' with respect to small perturbations, $x^n$.}) of the so-called ``butterfly catastrophe.''  As before, we will give the small generalization of the standard analysis of $W(x)$ to $V(\sigma,\bbox{\pi})$ and investigate the values of $R$ on the curves which separate the intrinsically different regions in the $(t, v, w)$ space.

For orientation, first consider the chiral symmetric potential, i.e., $w=0.$  The critical points of $W(x)$ are found from the roots of $W'(x)=0$. The five roots are at
\begin{equation}
x=0,\qquad x^2=\frac{1}{3}[-t\pm\sqrt{t^2-3v}].
\end{equation}
Consider the situation for different fixed values of t.

First take $t=0.$ There is a distinction between the $v>0$ and $v<0$ cases.  For $v>0$ there is only a single real root at $x=0$, which gives a minimum of $W(x)$.  This corresponds to a global minimum of $V(\sigma,\bbox{\pi})$ at the origin so the theory is in the Wigner-Weyl mode $(R=1)$.  For $v<0$ the real root at $x=0$ is a relative maximum, while the roots at $\pm(\frac{-v}{3})^{1/4}$ give two degenerate minima.  Translating to $V(\sigma,\pi_1)$ gives the Mexican hat shape and the Nambu-Goldstone mode characterized by $R=\infty.$

The same picture exists for $t>0$.  For $v>0$, $V(\sigma,\bbox{\pi})$ has just a minimum at the origin, so the theory is in the Wigner-Weyl mode $(R=1)$.  For $v<0$, $W(x)$ has a maximum at the origin and two degenerate minima at $x=\pm\frac{1}{\sqrt{3}}[-t+\sqrt{t^2-3v}]^{1/2}$.  This yields a Mexican hat shape for $V(\sigma,\pi_1)$, the Nambu-Goldstone mode, and $R=\infty$.

Unlike the above, the $t<0$ situation is quite different from the fourth order potential case.  Using (3.7) we see that for $v>t^2/3$, $W(x)$ has only a single real root at the origin.  This corresponds to a symmetrical ``bowl''-shaped $V(\sigma,\pi_1)$ so the theory is in the Wigner-Weyl mode $(R=1)$.  When $v<0$, $W(x)$ has a maximum at the origin and two degenerate minima at $x=\pm\frac{1}{\sqrt{3}}[-t+\sqrt{t^2-3v}]^{1/2}$.  This yields a Mexican hat shape for $V(\sigma,\pi_1)$ and $R=\infty$.  In the intermediate region $0<v<t^2/3$, we learn with the help of (3.7) that $W(x)$ has a local minimum at $x=0$, degenerate local maxima at $x=\pm\frac{1}{\sqrt{3}}[-t-\sqrt{t^2-3v}]^{1/2}$ and degenerate local minima at $x=\pm\frac{1}{\sqrt{3}}[-t+\sqrt{t^2-3v}]^{1/2}$.  The picture for $V(\sigma,\pi_1)$ is that of a central pond, ringed by mountains, which are in turn surrounded by a moat (Fig. 5).  If the pond is deeper than the moat the theory will be in the Wigner-Weyl  mode $(R=1)$, while if the moat is deeper than the pond the theory will be in the the Nambu-Goldstone mode $(R=\infty)$.  Now the pond and the moat will be equally deep if the locations of the outer minima of $W(x)$ are simultaneous solutions of $W(x)=W'(x)=0$.  This is seen to correspond to $x=\pm\sqrt{-t/2}$ and $v=t^2/4$.  For $0<v<t^2/4$ the moat is deeper while for $t^2/4<v<t^2/3$ the pond is deeper.  Putting these results together gives $R$ for the chirally-symmetric potential $(w=0)$ and negative $t$:
\begin{equation}
R= \left\{\begin{array}{l}  1,     \qquad (v>t^2/4)
			 \\ \infty, \qquad (v<t^2/4) \end{array}\right. \qquad (t<0).
\end{equation}

This may be contrasted with the $t\ge 0$ cases for the chirally symmetric potentials:
\begin{equation}
R= \left\{\begin{array}{l}  1,	\qquad v>0
			\\  \infty, \qquad v< 0 \end{array}\right. \qquad (t\ge 0).
\end{equation}

After this warmup we turn on chiral symmetry breaking by setting $w\ne 0.$  This is like adding a tilted plane to $V(\sigma,\pi_1)$ so that, for example, a circle of degenerate minima gets converted into a single minimum and a saddle point.  The discussion above suggests that it is sensible to analyze the problem in the $v-w$ plane for various fixed values of $t$.  Since $t$ is the coefficient of the fourth order term which was constrained by stability to be positive in the potential function of section II, we see at once that only positive non-zero values of $t$ correspond to structurally similar generalizations of the fourth order potential case.  Negative values of $t$ give rise to new possibilities.  Now, as discussed around eqs. (2.14) and (2.15), the locus of points where the number of minima of $W(x)$ changes is the simultaneous solution of $W'(x)=W^{\prime\prime}(x)=0.$.  This yields, for fixed $t$, the parametric relation \cite{catastrophe} between $v$ and $w$:
\begin{equation}
\begin{array}{l} v=-15x^4-6tx^2, \\
		 w=24x^5+8tx^3.		\end{array}
\end{equation}
Here the parameter $x$ corresponds to an inflection point of the function $W(x)$.

First consider the case where $t=0$.  From (3.10) we get the simple cusp,
\begin{equation}
v=-\frac{15}{(24)^{4/5}}w^{4/5},
\end{equation}
which may be contrasted with (2.16) [noting $v\to u$ and $w\to v$].  This is plotted in Fig. 6.  The physical picture is the same as that for the fourth order potential case:  $V(\sigma,\pi_1)$ has just a single global minimum above the cusp and a global minimum plus a relative maximum (and also a saddle point) below the cusp.  We already know $R$ on the $v$ axis from (3.9).  Setting $t=v=0$ in (3.5) immediately shows that $R=5$ along the $w$ axis.  It is more complicated to find $R$ on the cusp lines themselves;  in Appendix A, we show that a procedure similar to that used in (2.18)-(2.20) yields the constant value,
\begin{equation}  
\begin{array}{l} R=-\frac{5}{4}z(z^4-1), \\
	z\equiv -k^{1/3}+\frac{5}{9k^{1/3}}-\frac{2}{3},\ k\equiv \frac{5}{9}(\sqrt{6}+\frac{7}{3}). \end{array} .
\end{equation}
Numerically $R\approx 13.253.$  The various $R$ values are also displayed in Fig. 6.  The differences between the numerical values of $R$ here and in the case of the fourth order potential are due to the fact that the fourth order term is missing for $t=0$ so the sixth order term plays an important role.

The smooth generalization of the fourth order case of primary interest is to the $t>0$ situation.  In Fig. 7 we show the plot of (3.10) for $t=1$.  This is again a cusp which separates the potential into two intrinsically different regions.  Above the cusp $V(\sigma,\pi_1)$ has just one global minimum (generalized Wigner-Weyl region) while below the cusp (generalized Nambu-Goldstone region) $V(\sigma, \pi_1)$ has a global minimum together with a relative maximum (plus a saddle point).  It is amusing to notice that for large values of the parameter $|x|$ (3.10) reduces to the cusp (3.11) again while for small values of $|x|$ it reduces to a cusp of the type (2.16) where, however, the ``control parameters'' have been rescaled.  We have also indicated in Fig. 7 the $R$ values along the $v$ axis as obtained in (3.9) and the asymptotic tendencies of the $R$ values (for any $t>0$) along the $w$ axis and along the cusp lines as obtained numerically.  In each case, $R$ goes monotonically between the asymptotic limits.  The asymptotic values can be very easily understood.  Referring to (3.6), we find that when, for fixed $t$, $|v|$ and/or $|w|$ become large we can neglect the $tx^4$ term compared to the others.  Hence, we are essentially in the $t=0$ regime already discussed.  Indeed, if we neglect the central region, the asymptotic $R$ labelling in Fig. 7 reduces to that of Fig. 6.  On the other hand, when $|v|$ and $|w|$ become small we may neglect the $x^6$ term in (3.6) so that we have (taking $t>0$)
\begin{equation}
\frac{W(x)}{tA} \to x^4+\frac{v}{t}x^2+\frac{w}{t}x.
\end{equation} 
This is the fourth order potential case with rescaled $v$ and $w$.  However, the values of $R$ along the cusp lines and along the horizontal axis are independent of $v$ and $w$.  Indeed we see that the $R$ labelling in the very central part of Fig. 7 reduces to that of Fig. 3.  Thus, even though $R$ now varies along the ``bifurcation set'' dividing the two kinds of potentials, this variation is rather small and smooth and does not change the qualitative picture!  At least, that is the situation as long as $t$ remains non-negative.

The $t$ negative case gives a rather different picture.  Thus it would be an expression of a different kind of physics, most likely the extension of the model to conditions of non-zero temperature or other thermodynamic variable.  Actually, it is related to the Ginzburg-Landau model for a first-order phase transition.  For negative $t$, the parametric relation between $v$ and $w$ given by (3.10) leads \cite{catastrophe} to a ``butterfly'' shape with three cusps.  This is illustrated in Fig. 8, in which the locations of the key features have been indicated.  The arrows on the lines of the bifurcation set indicate how it is traversed as the parameter $x$ runs from $-\infty$ to $+\infty$.  The origin corresponds to $x=0.$  As before, the entire picture is symmetric on reflection about the $v$ axis as a result of the property $W(x; t, v, w) = W(-x; t, v, -w)$.  For orientation, notice that as $|t|\to 0$, the butterfly structure collapses to the origin and we are left with Fig. 6 once more.  From this point of view, the butterfly is a baroque embellishment of the by now familiar cusp lines.

If we look in the region below the butterfly, between the two main cusp lines we find that, as before, $W(x)$ has two local minima.  Above the butterfly, as before, $W(x)$ has just a single global minimum.  The new topography is within the butterfly.  In the central diamond-shaped region $W(x)$ has three local minima, while in the two triangular-shaped regions $W(x)$ has two local minima.  Now consider the translation of these results to the desired chiral potential $V(\sigma,\pi_1)$.  Refer to Fig. 9 which, to avoid clutter, shows just the left half of the symmetric picture.  As before, region A contains the global minimum (which becomes a degenerate ring on the $v$ axis) plus a relative maximum.  Similarly, region D contains just the global minimum.  Inside the butterfly, region B generally contains potentials having two relative minima separated by a saddle point in addition to a relative maximum and one more saddle point.  (On the $v$ axis this becomes Fig. 5).  Still inside the butterfly, the region C contains two local minima separated by a saddle point.

Notice that regions B and C each contain potentials with two relative minima.  The dashed line which runs from the point $v=t^2/4$ on the $v$ axis to the tip of the upper cusp is the locus of potentials for which the two minima are equally deep (``Maxwell set'').  As this line is crossed, the minimum which had been deepest becomes the shallower one and vice-versa.  We take the stable ground state to ``sit'' at the deepest minimum.
 
Next, let us study the values of $R$ on the axes and on the boundary lines of Fig. 9.  The values on the $v$ axis were already obtained in (3.8).  On the $w$ axis, for large $|w|$, $W(x)$ will have only one real root.  Using (3.6) with $v=0$ it is easy to see that its location $|x| \to (\frac{|w|}{6})^{1/5}$.  Since this gets large for large $|w|$, (3.5) shows that $R\to 5$.  The situation is different as one approaches the origin along the $w$ axis. Then for negative $t$, the denominator of (3.5) develops a pole and it is better to evaluate $R$ as the ratio of (3.3) to $m^2_\pi=-wA/\langle\sigma\rangle$.  When $|w|$ is small we find $|\langle\sigma\rangle|\approx \sqrt{-\frac{2t}{3}}$ and the leading term in an expansion for $R$ to be 
\begin{equation}
R=\frac{c(t)}{|w|},\qquad c(t)=\frac{16t^2}{3}\sqrt{-\frac{2t}{3}} \\
\qquad (v=0, w\to 0).
\end{equation}
The values of $R$ on the axes are indicated in Fig. 9.  As we go from infinity to the origin along the $w$ axis, $R$ increases monotonically from 5 to $\infty$ where it smoothly matches the values coming from the $v$ axis.

Now consider the values of $R$ on the curves separating the intrinsically different kinds of potential functions.  First it is amusing to note that $R=0$ at the tip of the upper cusp, corresponding to a massless sigma and a massive pion.  The derivation is given in Appendix A.  We investigated the remaining behavior numerically.  As expected, $R$ approaches $\approx 13.25$ on the main cusp line for large $|w|$ and negative $v$.  As one returns to the butterfly region $R$ increases monotonically to about 30.5 when it reaches the Maxwell set (dashed line in Fig. 9).  At this point it suddenly drops to about 0.95 and then slowly rises to 1 when it reaches the point $v=t^2/3$ on the $v$ axis.  Note that $R$ is two-valued on the Maxwell set.  On the line from the origin to the tip of the upper cusp, $R$ decreases monotonically from infinity to zero.  It has the same general behavior on the ``lower lip'' of the Maxwell set which runs from $v=t^2/4$ on the $v$ axis to the upper cusp tip.  On the line from $v=t^2/3$ on the $v$ axis to the upper cusp tip, R decreases monotonically from 1 to 0.  Finally, $R$ has the same behavior as one goes on the ``upper lip'' of the Maxwell set from $v=t^2/4$ on the $v$ axis to the upper cusp tip.

It is useful to exploit the fact that the function $W(x; t,v,w)$ in (3.6) is invariant, up to an irrelevant overall factor, under the scaling transformation \cite{catastrophe}:
\begin{equation}
x'=\lambda x,\quad t'=\lambda^2 t,\quad v'=\lambda^4 v, \quad w'=\lambda^5 w.
\end{equation}
As a consequence if we choose $t=-1$ we have $v'=t'^2 v$ and $w'=(-t)^{5/2}w.$  This means that if the butterfly curves corresponding to $t=-1$ are plotted we may use the identical curves for $any$ negative value of $t$ simply by relabelling the $v$ and $w$ axes.  Furthermore, (3.5) shows that $R$ is invariant under this transformation:
\begin{equation}
R(t,v,w) = R(t',v',w').
\end{equation}
Since we have just seen that there is one ``universal'' plot for all butterfly curves we may evaluate, for example, the two values of $R$ at the point where the main cusp line crosses the Maxwell set for $t=-1$ and be confident that the result will hold for any negative $t$.

Actually, far from the central butterfly region the pattern for $R$ reverts to the previous ones in the $t\ge 0$ cases.  This pattern is not very different from the fourth order potential case.  In fact, we may conclude that, as long as $t\ge 0$, there is not much difference between the fourth and sixth order potential description.  The $t<0$ region may be relevant for studying different ``phases'' of the system.  It is interesting to notice that that, unlike the fourth order potential functions, the region of R space where $0\le R<1$ is now allowed when $t<0$.  Such theories will occur in the region near the upper cusp tips, both inside and outside the butterfly wings.  The new allowed range for $R$ can be regarded as an indication of a qualitatively new behavior.  Thus, even though $R$ is not strictly constant on the curves separating different regions of the sixth order potential functions, it remains a useful indicator of the various different possibilities.

Actually, the $0\le R<1$ region might correspond to interesting new physics if the sixth order potential is taken to model low energy QCD with $t<0$ representing a thermodynamically exotic regime.  Then since $m(\sigma)<m(\pi)$ there would be different decay modes allowed.  The $\sigma$ would decay as $\sigma\to\gamma\gamma$ rather than $\sigma\to\pi\pi$.  More relevant for possible phenomenology would be new pion decay modes.  In addition to $\pi^0\to\gamma\gamma$, it would also be allowed to have $\pi^0\to\sigma\gamma\gamma$, etc.  Thus the photon energies would form a continuum.  Furthermore, in addition to $\pi^-\to\l^-{\overline\nu}_l$ we would have $\pi^-\to\sigma l^-{\overline\nu}_l$ etc.  Evidently there would be different experimental signatures for the new regime!

\section{Discussion}

We saw in section II that the fourth order potential functions could be divided into two types characterized by different ranges of $R$:  1) For $1\le R<9$ the potential $V(\sigma, \pi_1)$ had the shape of a simple bowl.  2)  For $R>9$ the potential had, in addition to one minimum value (which could be continuously degenerate), an extra bump (or relative maximum).  We speculated that the $R<9$ region represented a generalization of the Wigner-Weyl type physics in the sense that a theory in it could be accurately reached by perturbation starting from $R=1$.  The $R>9$ theories (generalized Nambu-Goldstone) could presumably be reached by perturbation starting from $R=\infty.$  Clearly, an interesting question is the dynamical one of what goes wrong if one tries to reach, say an $R>9$ theory from $R=1.$  This would probably depend on the space-time dimension and might be complicated to implement.  In any event, the classification presented has an evident ``kinematical'' significance.  There may be some connection for the case of the application to QCD with the work of the Orsay group \cite{Orsay}.

It might be worthwhile to consider generalizations to broken symmetry groups other than $SU(2)_L\times SU(2)_R$ [say $SU(N)_L\times SU(N)_R$] and to more complicated particle multiplets.  In section III we investigated the effect of using a sixth order rather than a fourth order potential.  Comparing Fig. 3 with the central part of Fig. 7 showed that, for parameters of the sixth order potential yielding just two regions, there was not much change.  The main difference is the existence of new regions, as in Fig. 9, for the sixth order case.  These are signaled by the opening of the range $0\le R<1$ and might correspond to QCD in a special situation.  Presumably a similar pattern continues for still higher order potential functions.    

\centerline{\bf Acknowledgements}

We are happy to thank Masayasu Harada, Francesco Sannino and Sachindeo Vaidya for helpful discussions.  This work was supported in part by the U.S. DOE, Contract No. DE-FG-02-ER40231.

\section*{Appendix A}

\setcounter{equation}{0}
\renewcommand{\theequation}{A.\arabic{equation}}

Here we give some details of the computations in the sixth order potential case.  On the bifurcation set (locus of points where the number of minima changes) a minimum and a maximum must have coalesced to an inflection point so that $W'(x)$ must have a double root.  We then parameterize:
\begin{equation}
\frac{1}{A}W'(x)=6(x-a)(x-b)^2(x^2+px+q),
\end{equation}
where $x=b$ is the location of the inflection point.  Comparing this with $W'(x)$ as obtained from (3.6) gives the identifications:
\begin{equation}
\begin{array}{l}
	p=a+2b,\qquad q=\frac{2b(a+b)^2}{a+2b},\\
	w=-6qab^2,\qquad v=3[-ab^2p+qb(2a+b)],\\
	t=-\frac{3}{2} \frac{a^3+4b^3+3ab^2+2a^2b}{a+2b}. \end{array}
\end{equation}
Now consider the $t=0$ case.  $W(x)$ has a region with one minimum and another with two minima.  Hence $W'(x)$ has at most three real roots.  On the bifurcation set we must identify $x=a$ as the location of the global minimum.  Taking the ratio of (3.3) to (3.4) gives
\begin{equation}
R=\frac{30a^4+2v}{(-w/a)}=\frac{30a^4-30b^4}{(-24b^5/a)}= -\frac{5}{4}(z^5-z),
\end{equation}
where $z=a/b$ and we set $v=-15b^4$ and $w=24b^5$ (see (3.10)).  To find $z$ we impose $t=0$ in (A2) to get
\begin{equation}
z^3+2z^2+3z+4=0.
\end{equation}
The formula for the real root of this cubic equation is finally given in (3.12).

It is interesting to consider the case $a=b$, where (A.1) has a triple root.  Then from (A.2) we read off
\begin{equation}
p=3a,\quad q=\frac{8}{3}a^2,\quad w=-16a^5, \quad v=15a^4, \quad t=-5a^2.
\end{equation}
Equivalently, the locations of these points in the $w-v$ plane are $(w,v)=(\pm \frac{16t^2}{25}\sqrt{-\frac{t}{5}}, \frac{3t^2}{5})$ which are seen from Fig. 8 to agree with the locations of the tips of the upper cusps in the $t<0$ situation.  Substituting $\langle\sigma\rangle=a$ as well as (A.5) into (3.5) then gives $R=0$ for these special points.

\centerline{\bf Figure Captions}

FIG. 1(a). ``Mexican hat'' potential ($\lambda>0, B=0$).

FIG. 1(b). ``Symmetrical bowl'' potential ($\lambda<0, B=0$).

FIG. 1(c).  Perturbed ``Mexican hat'' potential ($B\ne 0$).
  
FIG. 2.  Bifurcation set (cusp) for the fourth order potential.  The hatched region below the cusp describes $W(x)$ functions with two minima.

FIG. 3.  Variation of $R$ throughout the $uv$-plane.  $R=\infty$ on the negative $u$ axis and $R=1$ on the positive $u$ axis.

FIG. 4.  Plot of (2.22) for application of linear $\sigma$ model to low energy QCD.

FIG. 5.  Sixth order potential $V(\sigma,\pi_1$) with $w=0$ and $t<0$.  This example corresponds to the region $0<v<\frac{1}{3}t^2$.
 
FIG. 6.  Sixth order potential cusp for $t=0$ and corresponding behavior of R.  $w$ is plotted along the horizontal axis and $v$ along the vertical axis.

FIG. 7.  Sixth order potential cusp for $t>0$ and corresponding behavior of R.

FIG. 8.  Sixth order potential bifurcation set for $t<0$.

FIG. 9.  Different topographic regions for the sixth order potential case when $t<0$.  The dashed line is a sketch of the ``Maxwell set''.  Also shown are the $R$ values along the $w$ and $v$ axes.  Note that the quantity $c$ is defined in Eq. (3.14).

\begin{center}
\epsfxsize=300pt
\epsfysize=250pt
\epsfbox{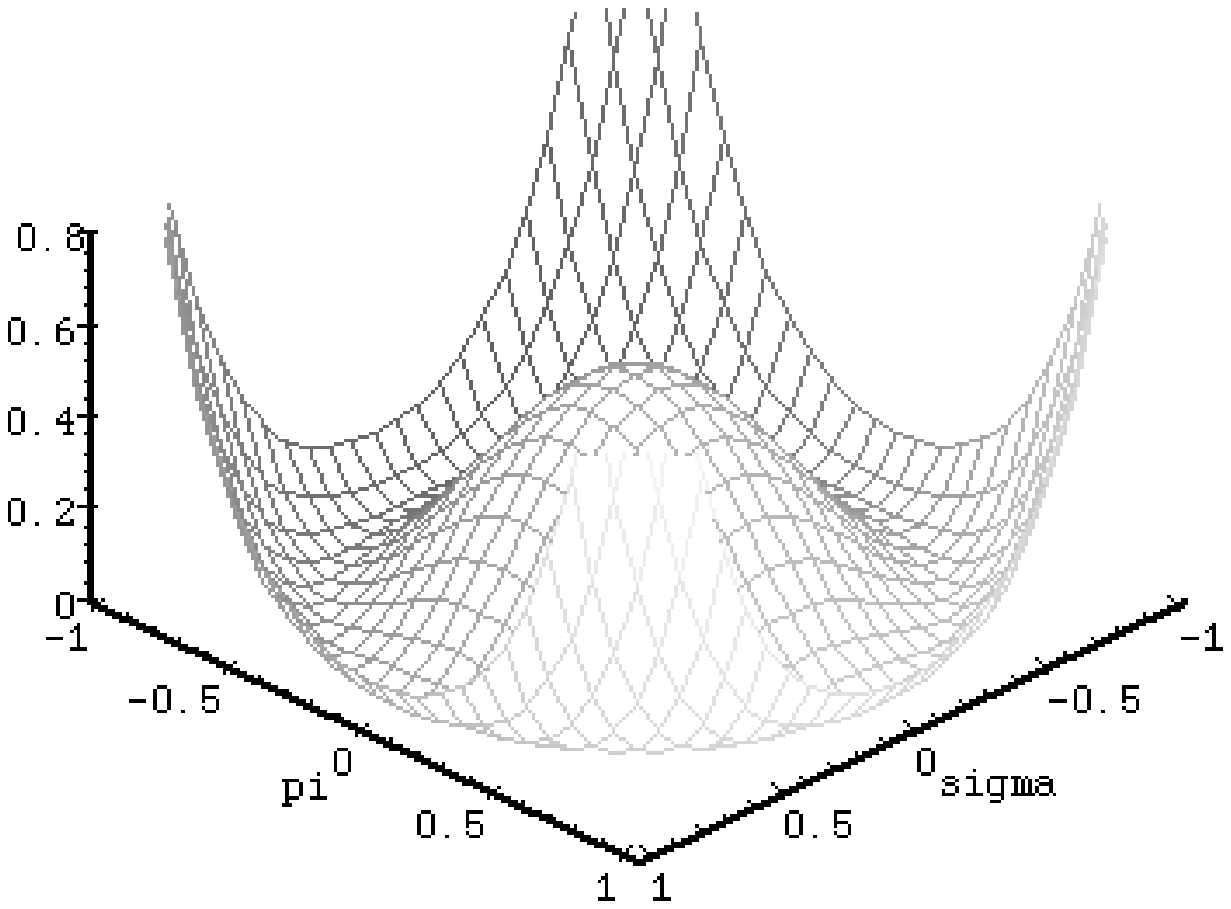}
\begin{itemize}
\item[Fig. 1a]

\end{itemize}
\end{center}

\begin{center}
\epsfxsize=300pt
\epsfysize=250pt
\epsfbox{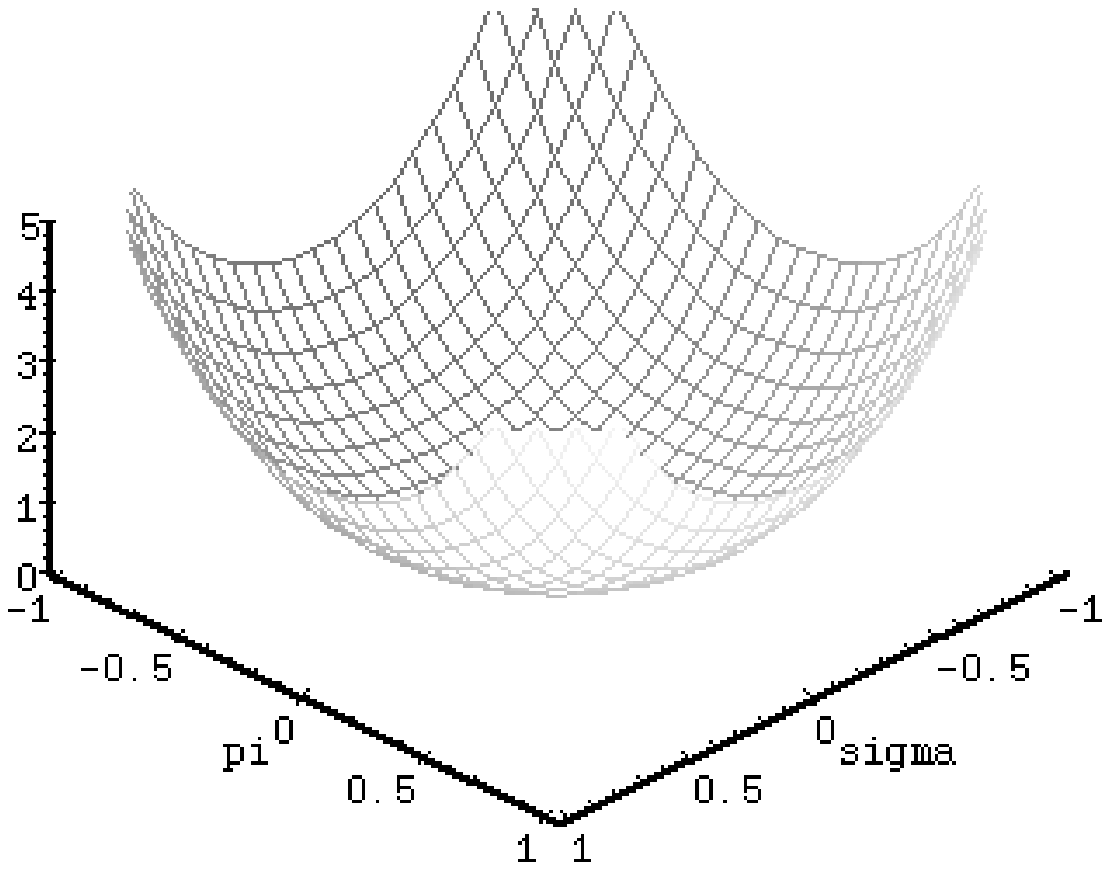}
\begin{itemize}
\item[Fig. 1b]
  
\end{itemize}
\end{center}

\begin{center}
\epsfxsize=300pt
\epsfysize=250pt
\epsfbox{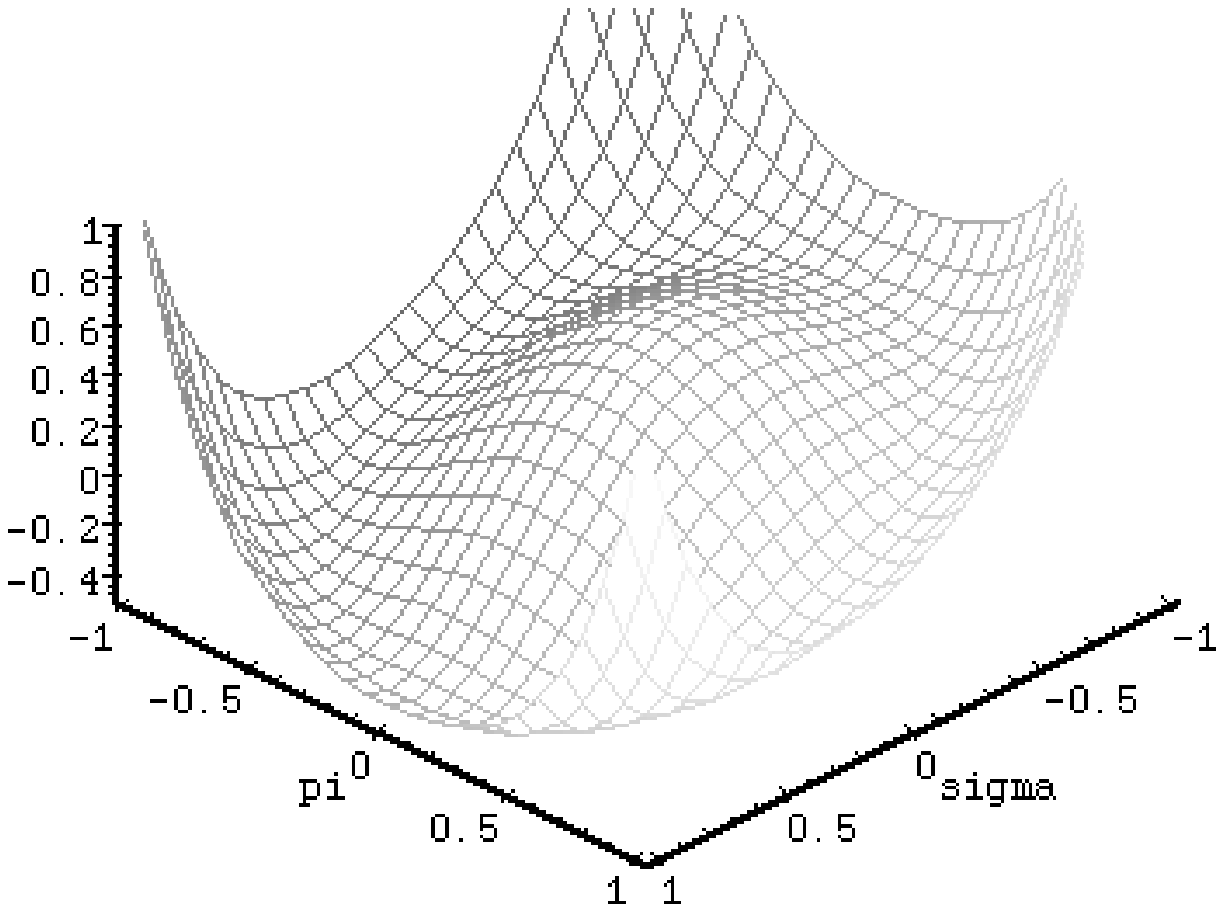}
\begin{itemize}
\item[Fig. 1c]
 
\end{itemize}
\end{center}

\begin{center}
\epsfxsize=300pt
\epsfysize=250pt
\epsfbox{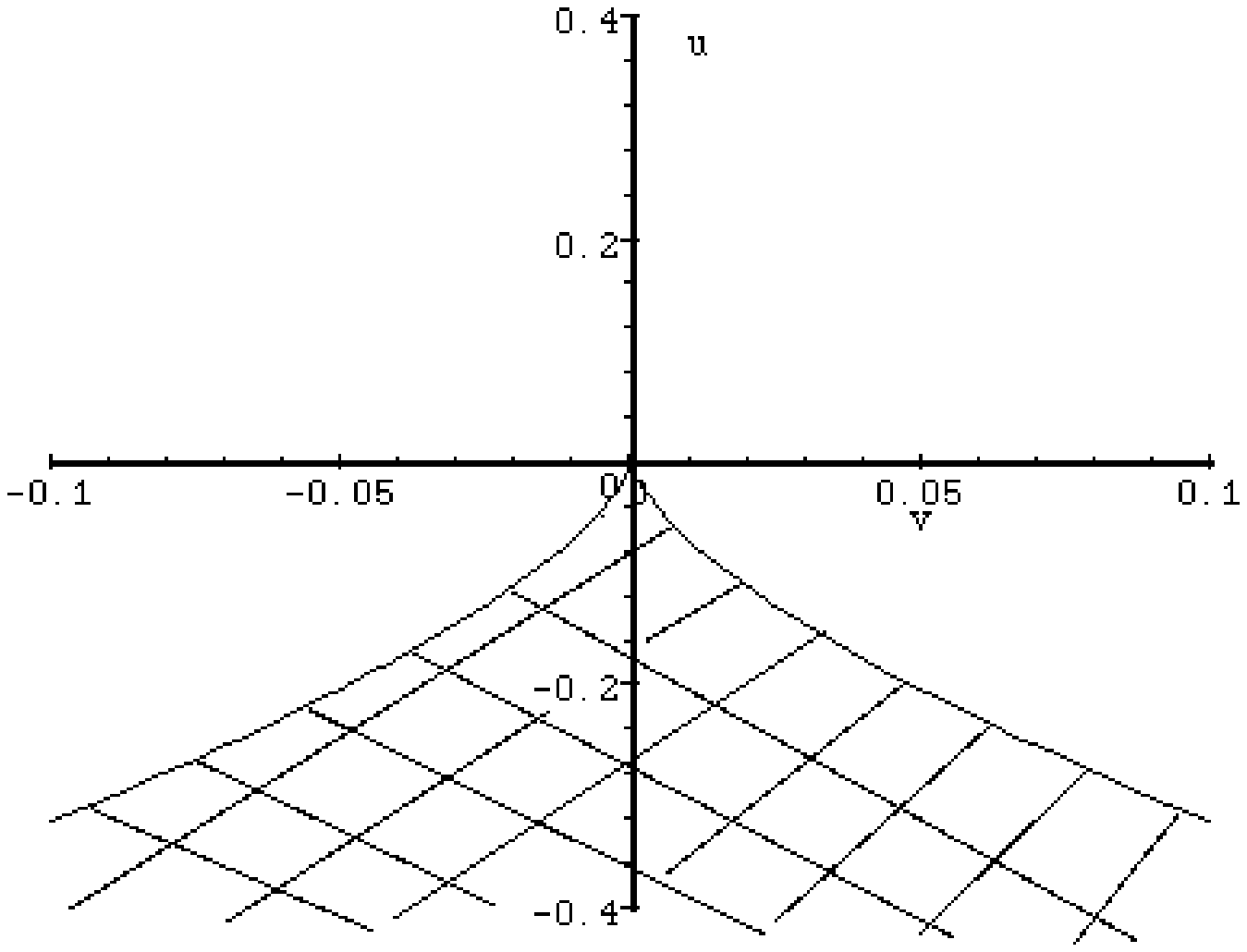}
\begin{itemize}
\item[Fig. 2]
  
\end{itemize}
\end{center}

\begin{center}
\epsfxsize=300pt
\epsfysize=250pt
\epsfbox{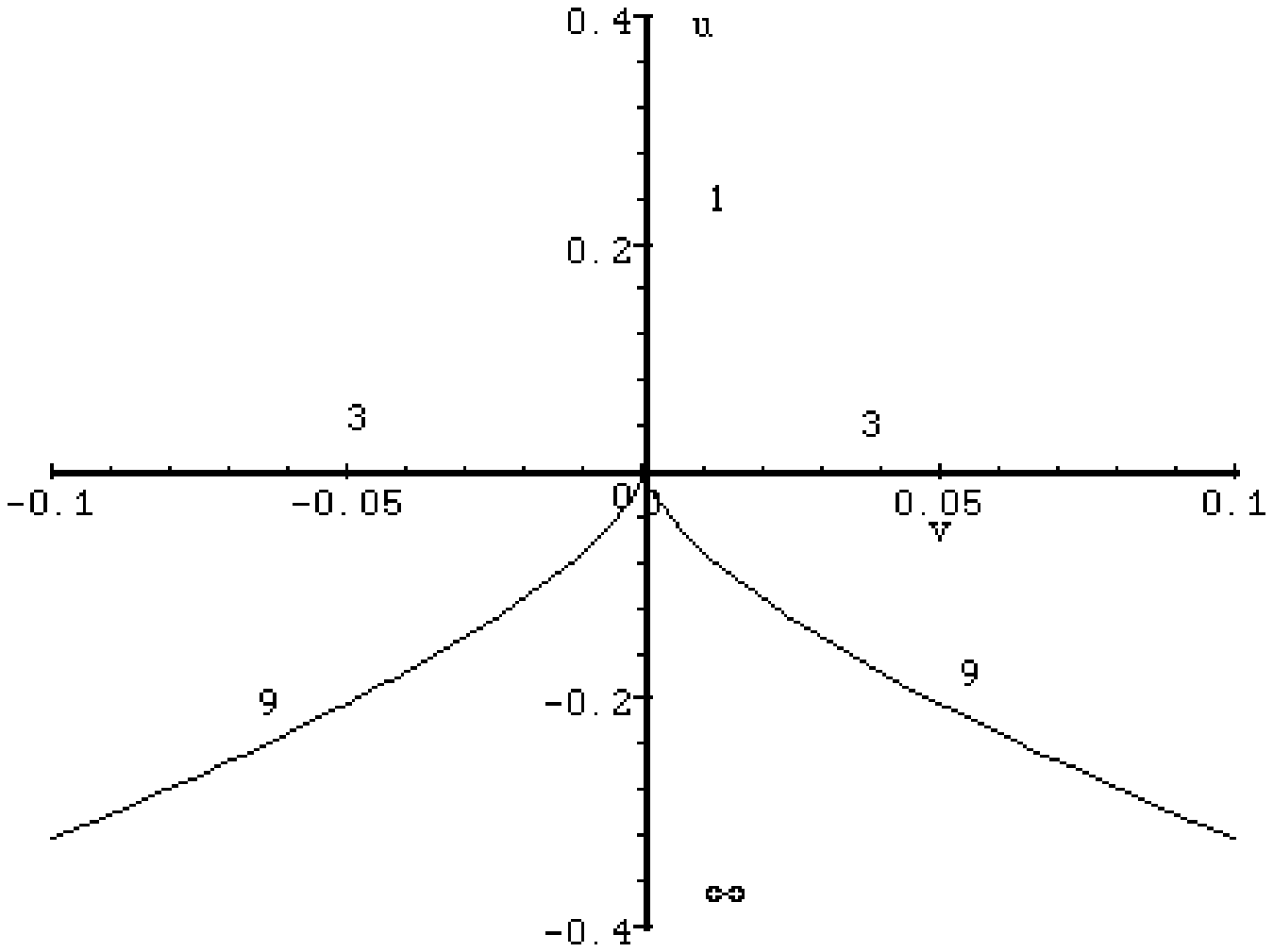}
\begin{itemize}
\item[Fig. 3]
  
\end{itemize}
\end{center}

\begin{center}
\epsfxsize=300pt
\epsfysize=250pt
\epsfbox{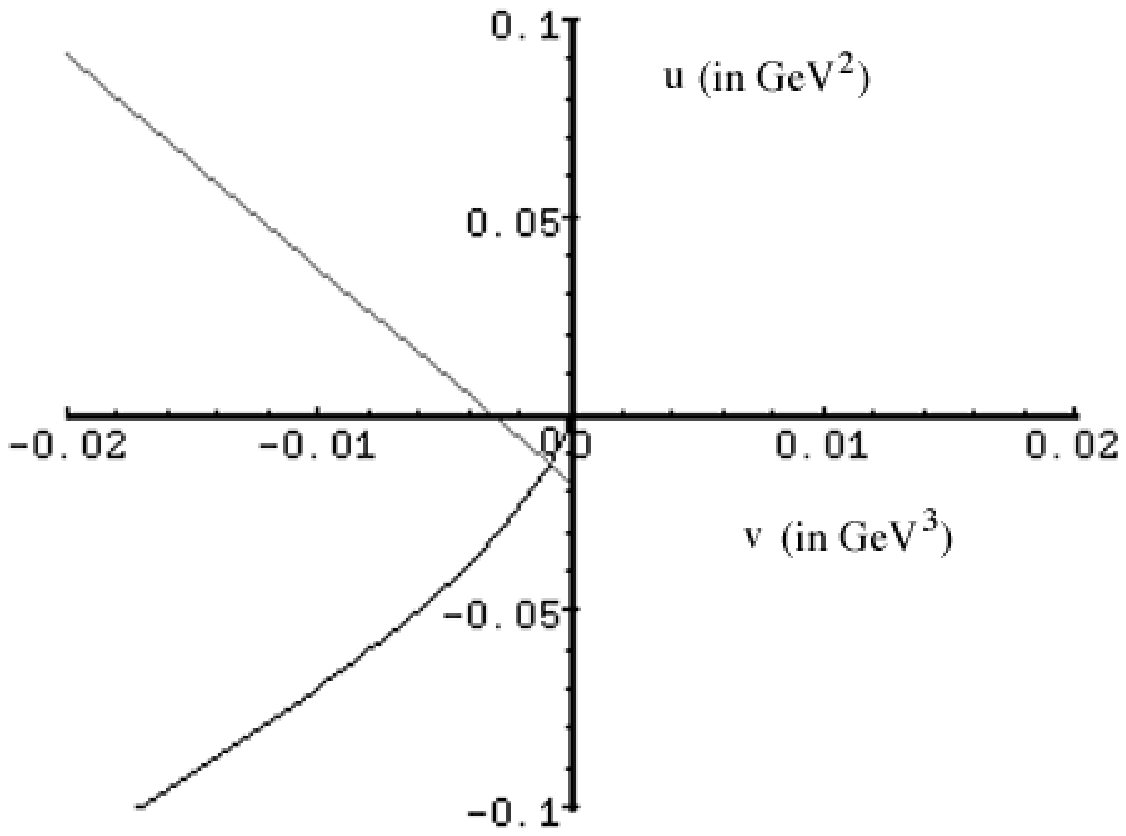}
\begin{itemize}
\item[Fig. 4]
  
\end{itemize}
\end{center}

\begin{center}
\epsfxsize=300pt
\epsfysize=250pt
\epsfbox{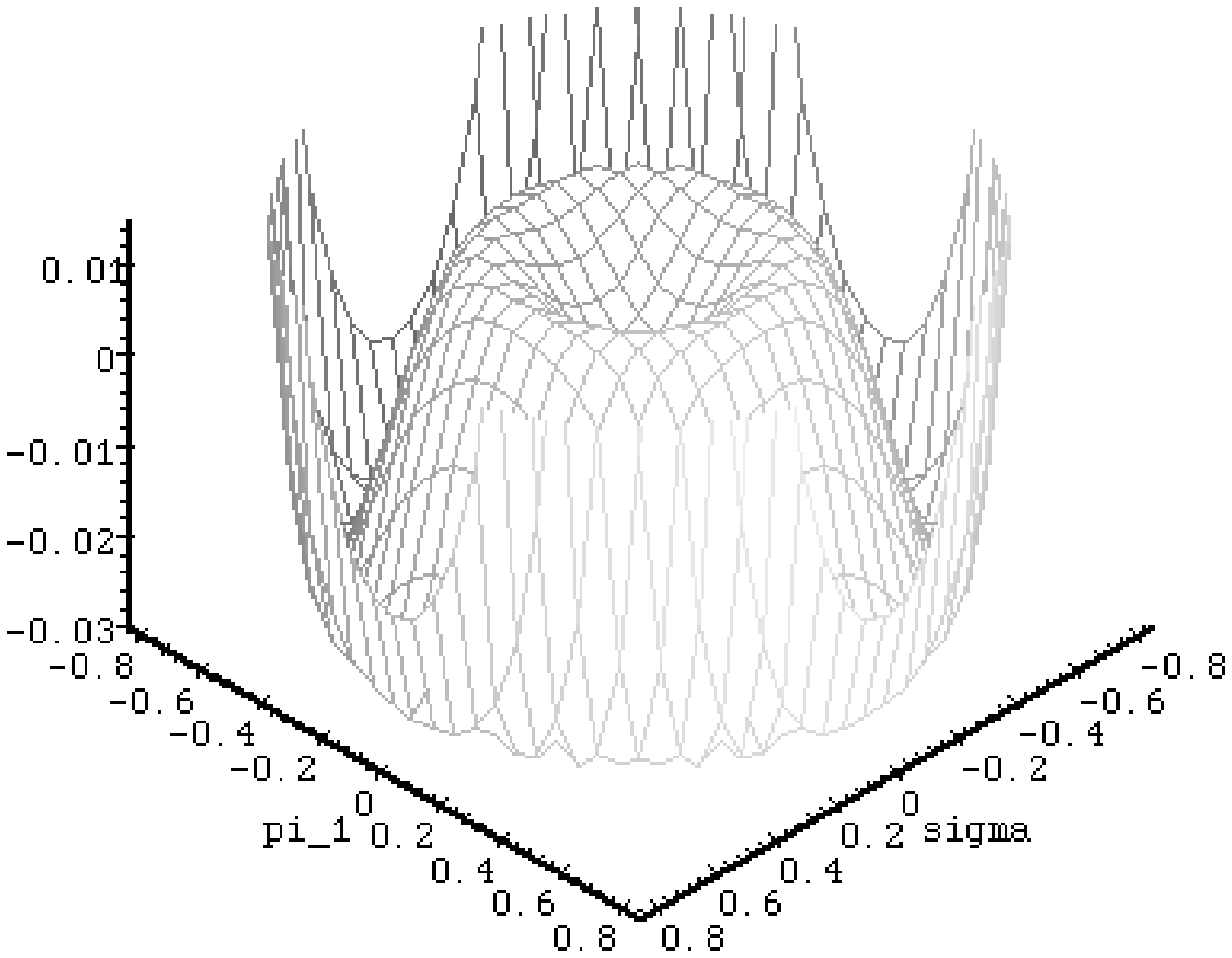}
\begin{itemize}
\item[Fig. 5]
  
\end{itemize}
\end{center}

\begin{center}
\epsfxsize=300pt
\epsfysize=250pt
\epsfbox{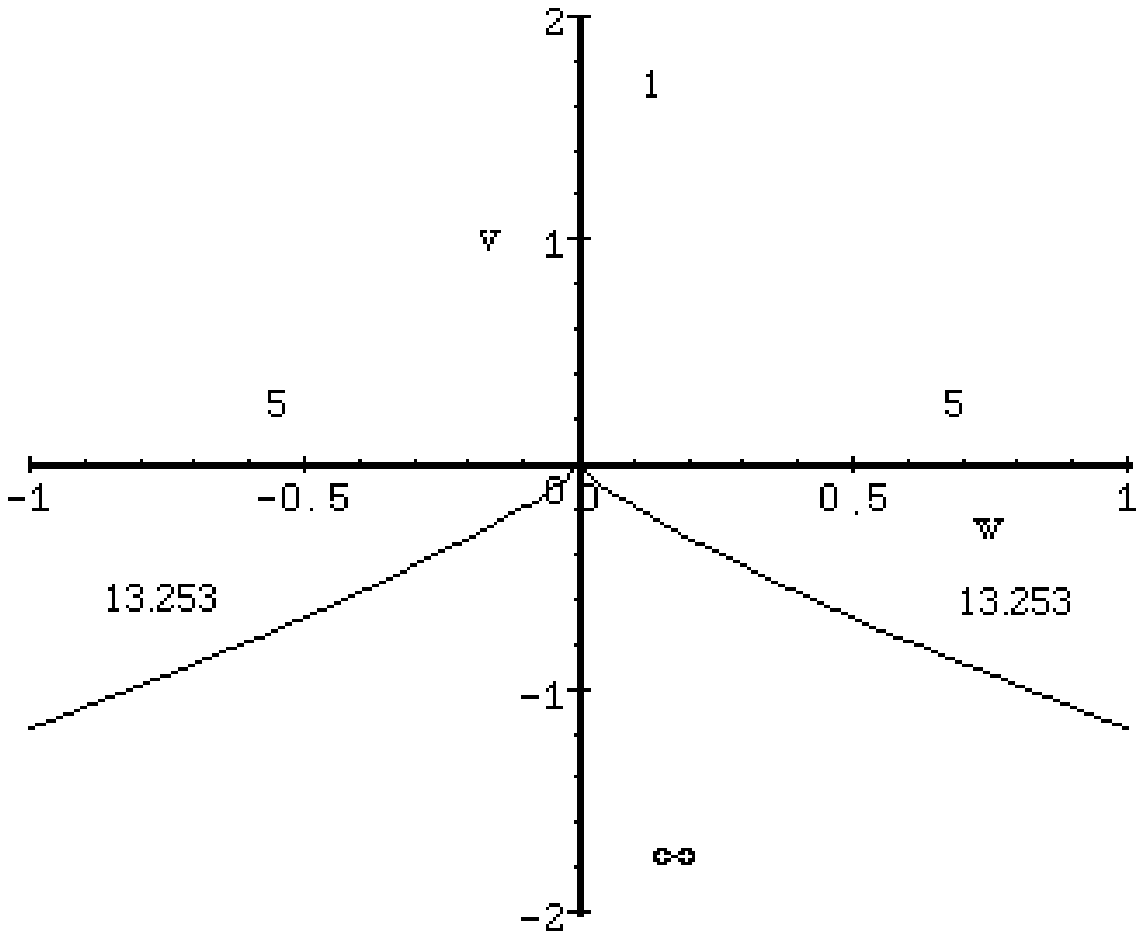}
\begin{itemize}
\item[Fig. 6]
  
\end{itemize}
\end{center}

\begin{center}
\epsfxsize=300pt
\epsfysize=250pt
\epsfbox{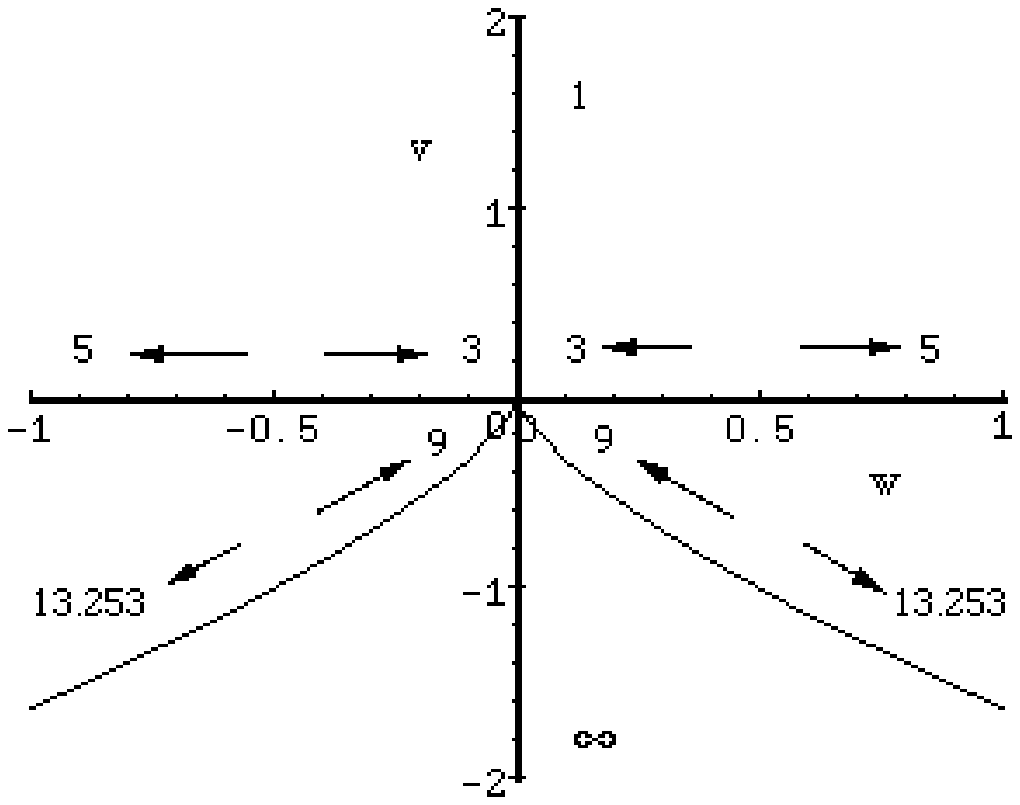}
\begin{itemize}
\item[Fig. 7]
  
\end{itemize}
\end{center}

\begin{center}
\epsfxsize=300pt
\epsfysize=250pt
\epsfbox{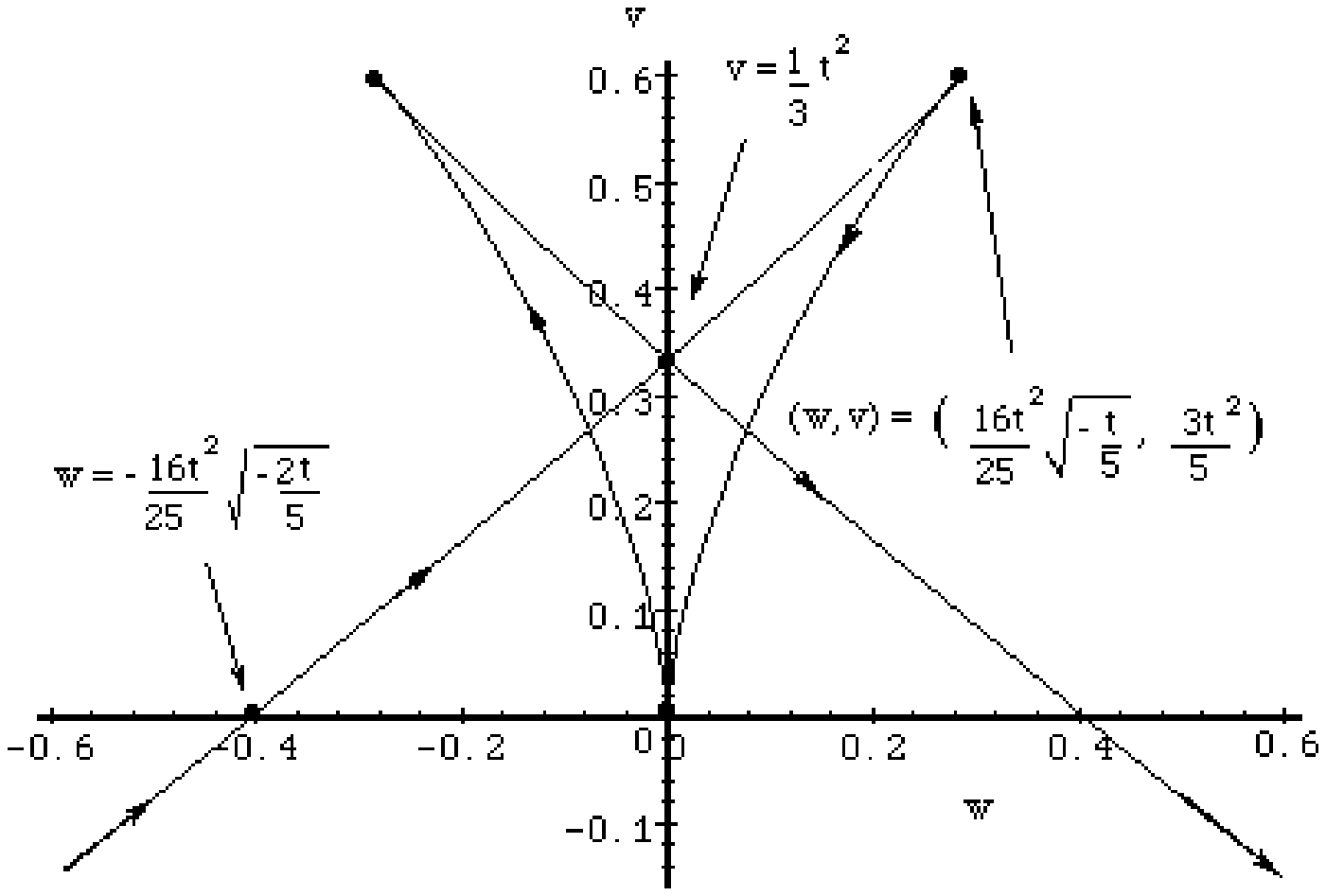}
\begin{itemize}
\item[Fig. 8]
  
\end{itemize}
\end{center}

\begin{center}
\epsfxsize=300pt
\epsfysize=250pt
\epsfbox{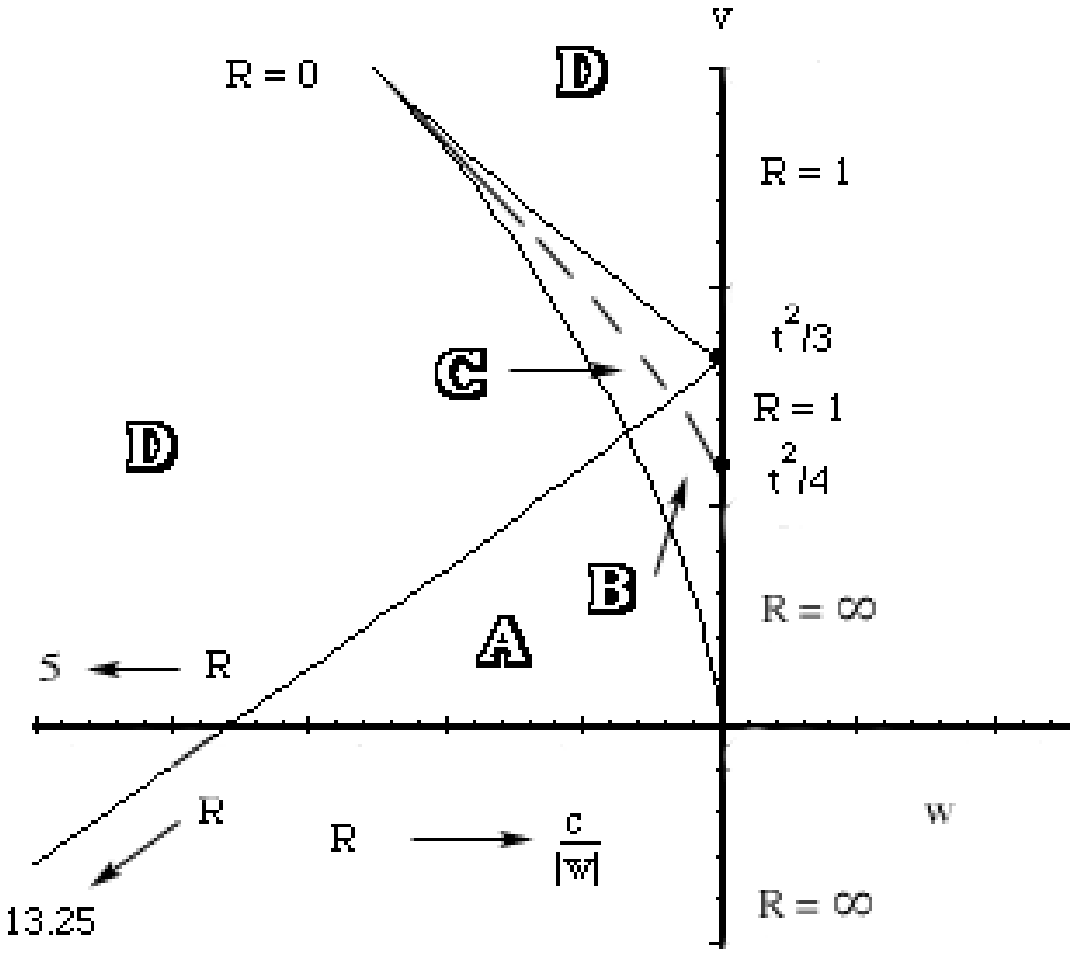}
\begin{itemize}
\item[Fig. 9]
  
\end{itemize}
\end{center}

\end{document}